# Electron-polaron dichotomy of charge carriers in perovskite oxides

M.-A. Husanu[1,2 ✉], L. Vistoli[3], C. Verdi[4,5], A. Sander[3], V. Garcia[3], J. Rault[6], F. Bisti[1], L. L. Lev[7], T. Schmitt[1], F. Giustino[4,8,9], A. S. Mishchenko[10,11], M. Bibes[3] & V. N. Strocov[1 ✉]

Many transition metal oxides (TMOs) are Mott insulators due to strong Coulomb repulsion between electrons, and exhibit metal-insulator transitions (MITs) whose mechanisms are not always fully understood. Unlike most TMOs, minute doping in $CaMnO_3$ induces a metallic state without any structural transformations. This material is thus an ideal platform to explore band formation through the MIT. Here, we use angle-resolved photoemission spectroscopy to visualize how electrons delocalize and couple to phonons in $CaMnO_3$. We show the development of a Fermi surface where mobile electrons coexist with heavier carriers, strongly coupled polarons. The latter originate from a boost of the electron-phonon interaction (EPI). This finding brings to light the role that the EPI can play in MITs even caused by purely electronic mechanisms. Our discovery of the EPI-induced dichotomy of the charge carriers explains the transport response of Ce-doped $CaMnO_3$ and suggests strategies to engineer quantum matter from TMOs.

[1] Swiss Light Source, Paul Scherrer Institute, 5232 Villigen-PSI, Switzerland. [2] National Institute of Materials Physics, Atomistilor 405A, 077125 Magurele, Romania. [3] Unité Mixte de Physique, CNRS, Thales, Université Paris-Sud, Université Paris-Saclay, 91767 Palaiseau, France. [4] Department of Materials, University of Oxford, Parks Road, Oxford OX1 3PH, UK. [5] Faculty of Physics and Center for Computational Materials Science, University of Vienna, 1090 Vienna, Austria. [6] Synchrotron SOLEIL, L'Orme des Merisiers Saint-Aubin, BP 48, 91192 Gif-sur-Yvette, France. [7] Moscow Institute of Physics and Technology, 9 Institutskiy lane, 141700 Dolgoprudny, Russia. [8] Oden Institute for Computational Engineering and Sciences, University of Texas at Austin, Austin, TX 78712, USA. [9] Department of Physics, The University of Texas at Austin, Austin, TX 78712, USA. [10] RIKEN Center for Emergent Matter Science (CEMS), 2-1 Hirosawa, Wako, 351-0198 Saitama, Japan. [11] National Research Center "Kurchatov Institute", 123182 Moscow, Russia. ✉email: ahusanu@infim.ro; vladimir.strocov@psi.ch

A hallmark of correlated transition metal oxides (TMOs) is a prominent cross-coupling between the electronic charge, spin, orbital, and lattice degrees of freedom. The concomitant rich physics emerges as a promising perspective for novel electronic devices for transistors[1–3], energy harvesting[4–6], or information storage and processing[7–9] including quantum computing. A promising strategy to develop new functionalities is to work with materials at the verge of a phase transition. For Mott insulators, this defines the field of Mottronics. This approach is especially suitable for the functionalization of giant conductivity variations across the metal-insulator transition (MIT)[10,11], with the change driven by various mechanisms such as structural distortions[12], substrate-induced strain[13], thickness[14,15], and bias[11], as well as for the switching of the spin order between different magnetic ground states[9,16].

$CaMnO_3$ belongs to a vast class of perovskite oxides, which are insulating due to strong electronic Coulomb repulsion. It relaxes into an orthorhombically distorted structure due to the strain induced by the different ionic radii of the Ca and Mn cations, consistently with its tolerance factor $t = (r_{Ca} + r_O)/2(r_{Mn} + r_O) = 0.74$ in the range of orthorhombic distortions[17], while electron-spin coupling stabilizes a G-type antiferromagnetic (AFM) order[18,19]. Unlike in many perovskite oxides[15,18,20], minute doping of $CaMnO_3$ can induce a sharp transition to a metallic state. The critical electron concentration $n_e$ for metallicity is in the low $10^{20}$ cm$^{-3}$ range[21], which exceeds the values typical of classical semiconductors by two orders of magnitude but does not introduce any significant disorder or structural changes. $Ce^{4+}$ doping at the $Ca^{2+}$ site, bringing two additional electrons per formula unit, pushes the Fermi level $E_F$ into the conduction band. The consequent mixed $Mn^{3+}/Mn^{4+}$ valence state accounts for electronic properties and magnetic ground state driven by the interplay of electron itinerancy due to double exchange interaction with polaronic self-localization[22–25]. This compound thus appears as an ideal platform to explore the effects of electron doping on the interaction of electrons with collective degrees of freedom[22–25].

Here, by doping $CaMnO_3$ (CMO) with only 2 and 4% Ce (CCMO2 and CCMO4), we explore the nature of charge carriers emerging at the early stages of band filling inducing the transition from the insulating to a metallic state. We use for this purpose angle-resolved photoelectron spectroscopy (ARPES), which directly visualizes the electronic band structure and the one-electron spectral function $A(\mathbf{k},\omega)$, reflecting many-body effects of electron coupling with other electrons and bosonic excitations. Our results show that the doping of CMO forms a system of dichotomic charge carriers, where three-dimensional (3D) light electrons, weakly coupled to the lattice, coexist with quasi-two-dimensional (q2D) heavy strongly coupled polarons. The latter form due to a boost of the electron–phonon interaction (EPI) on the verge of MIT where the occupied bandwidth stays comparable with the phonon energy. As in CCMO the energy scale of all elementary electronic and magnetic excitations is well separated from that of phonons, it gives us a unique opportunity to single out the EPI-specific effect. We argue that the EPI-boost scenario is universal for the MITs in other oxides.

## Results

**Theoretical electronic structure overview.** We start with the analysis of the overall electronic structure of the parent CMO suggested by ab initio calculations (see also Supplementary Note 1). The relaxed orthorhombic structure of this material is shown in Fig. 1a together with the corresponding Brillouin zone (BZ) in comparison to that of the ideal cubic lattice in Fig. 1b. The calculations confirm that the spin ground state of CMO adopts the G-type AFM order[18,19,26]. The calculated density of states, projected on the orbital contribution of Mn for the two sublattices (Fig. 1c), indicates that the valence states of CMO derive mostly from the half-filled $t_{2g}$ orbitals while the empty conduction states are a combination of $e_g$ and $t_{2g}$ derived states. The calculated band structure as a function of binding energy $E_B = E - E_F$ and momentum $\mathbf{k}$ is shown in Fig. 1d, schematically depicting the $e_g$ and $t_{2g}$ orbitals. In this plot $E_F$ is rigidly shifted according to the doping level. Already the low Ce-doping in CCMO2 pushes $E_F$ above the conduction band (CB) minimum and starts populating the electronic states having predominantly the Mn $e_g$ $x^2 - y^2$ character (Fig. 1c). The increase of doping to 4% continues populating the CB-states. Figure 1e shows the theoretical Fermi surface (FS) unfolded to the cubic pseudo-cell BZ, which consists of 3D electron spheres around the Γ point formed by the $e_g$ $3z^2 - r^2$ and $x^2 - y^2$ orbitals, and q2D electron cylinders along the ΓX and ΓZ directions formed by the $e_g$ $x^2 - y^2$ orbitals (Supplementary Fig. 1). Such manifold is characteristic of perovskites[15,20]. The FS cuts in the ΓMX plane are shown in Fig. 1f for the two doping levels.

**Experimental electronic structure under doping.** We will now discuss the experimental electronic structure of CCMO at different doping levels. Our $Ca_{1-x}Ce_xMnO_3$ samples ($x = 2, 4\%$ nominal Ce concentrations) were 20 nm thin films grown by pulsed laser deposition on the $YAlO_3$ (001) substrates (for details see section Methods: Sample growth). Our experiment (see section Methods: SX-ARPES experiment) used soft-X-ray photons (SX-ARPES) with energy $h\nu$ of few hundreds of eV which enable sharp resolution of out-of-plane electron momentum $k_z$ and thereby full 3D momentum $\mathbf{k}$[27,28] as essential for the inherently 3D electronic structure of CCMO. The measurements were carried out at the SX-ARPES endstation[29] of the ADRESS beamline at the Swiss Light Source[30] using circularly polarized incident X-rays. Figure 2a, b present the experimental FS maps of CCMO2 and CCMO4 measured as a function of in-plane $(k_x,k_y)$-momentum near the $k_z = 8 \cdot (2\pi/a)$ plane ΓXM of the cubic pseudo-cell BZ (Fig. 1b). These measurements were performed with $h\nu = 643$ eV which pushes the probing depth in the 3.5 nm range[31] and increases signal from the Mn $3d$ states by a factor of 2–3 by their resonant excitation through the Mn $2p$ core levels (Supplementary Figs. 2, 3). Simultaneously, it sets $k_z$ close to the Γ-point of the BZ as evidenced by the experimental FS map as a function of $(k_x,k_z)$-momentum in Fig. 2c. The trajectory of the 643 eV energy in the $\mathbf{k}$-space is sketched for clarity in Fig. 2d. In agreement with the theoretical FS in Fig. 1e, the experiment distinctly shows the $e_g$ $3z^2 - r^2$ derived 3D electron spheres around the Γ-points and $e_g$ $x^2 - y^2$ derived q2D electron cylinders extending along ΓX. The experimental results are reproduced by the theoretical FS contours calculated for the 2% and 4% doping levels in Fig. 1f, although they slightly underestimate the Fermi vector $k_F$. The orthorhombic lattice distortion manifests itself in the experimental FS maps as replicas of the 3D spheres repeating every $(\pi/a, \pi/a)$ point of the cubic pseudo cell, and q2D cylinders every $(\pi/a, 0)$ and $(0, \pi/a)$ point. These replicas are analogous to the SX-ARPES data for the rhombohedrally distorted $La_{1-x}Sr_xMnO_3$[20].

The electron densities embedded in the 3D spheres and q2D cylinders, $n_{3D}$ and $n_{q2D}$ respectively are compiled in Table 1. These were obtained from the Luttinger volumes $V_L$ of the corresponding FS pockets determined from the experimental $k_F$ values. The latter are determined from the gradient of the ARPES intensity[32] integrated through the whole occupied bandwidth. This method was used because of its robustness to many-body effects and experimental resolution. Comparison of the experimental FS maps for CCMO2 and CCMO4 in Fig. 2a, b, respectively, shows clear increase of $V_L$ and thus electron density with doping for both 3D spheres and q2D cylinders. This is particularly clear for the cuts of the 3D spheres around the $\Gamma_{i,j}$ points which develop from small filled circles to larger open ones. We note that $n_{3D}$ increases much more with doping than $n_{q2D}$.

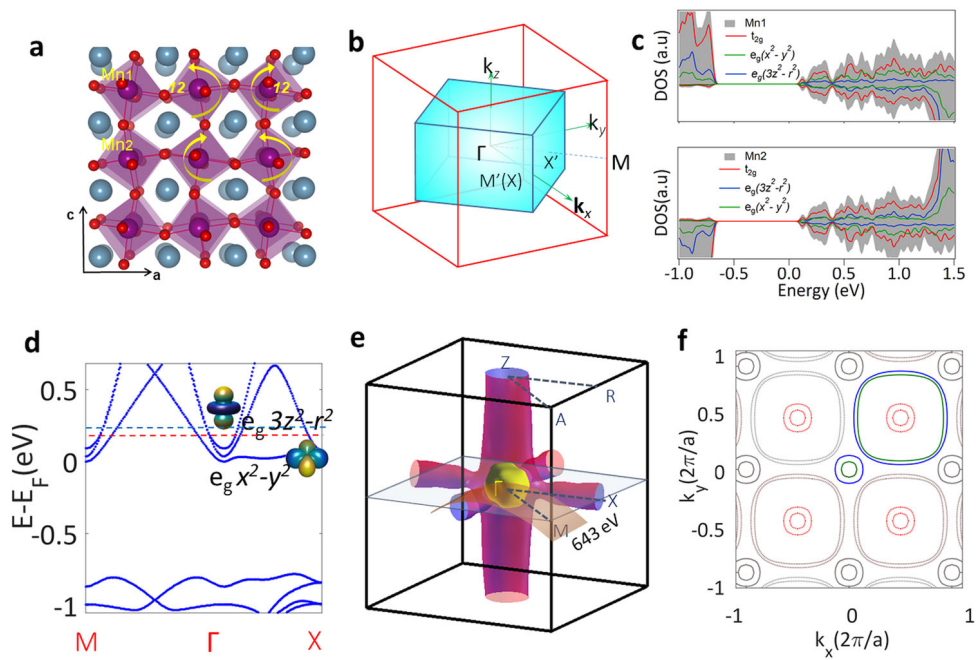

**Fig. 1 Theoretical electronic structure of CaCeMnO$_3$ (CCMO) upon doping. a** Density functional theory -relaxed orthorhombically distorted lattice of CaMnO$_3$ (CMO). **b** The Brillouin zone (BZ) of the distorted lattice inscribed into that of the undistorted cubic pseudo-cell. **c** Electronic density of states of CMO projected onto two Mn atoms with antiparallel spins. **d** Bandstructure and rigid shift of the Fermi energy $E_F$ corresponding to 2% (red) and 4% doping (blue). **e** Theoretical Fermi surface (FS) unfolded to the cubic pseudo-cell BZ. **f** Calculated FS cuts in the ΓMX plane for the 2% Ce-doped, CCMO2 (green lines) and 4% Ce-doped, CCMO4 (blue), where the doping increases the Luttinger volume.

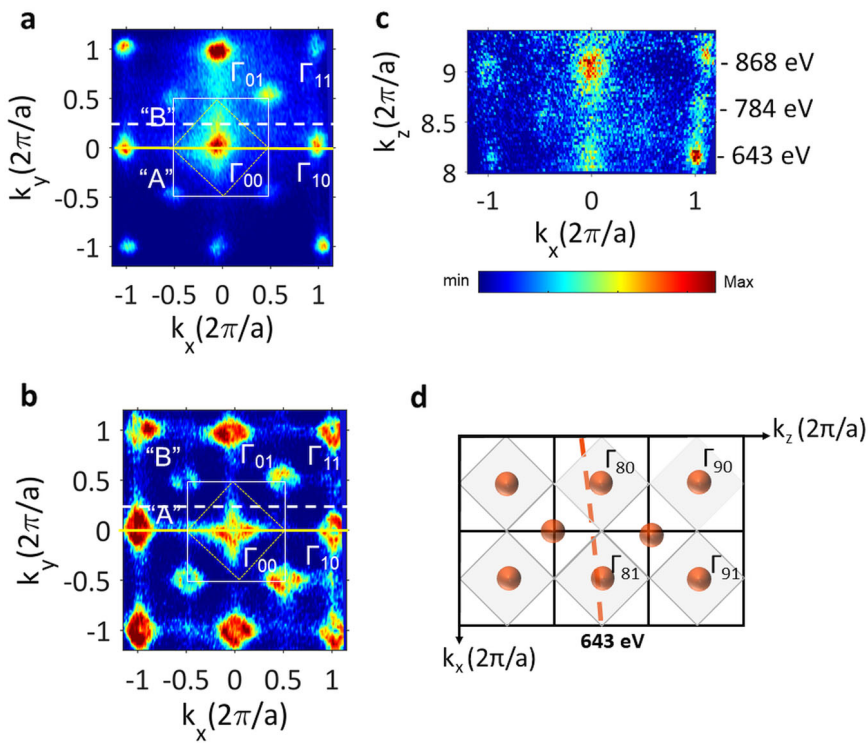

**Fig. 2 Experimental electronic structure of CaCeMnO$_3$ (CCMO). a, b** Out-of-plane Fermi surface (FS) maps in the ΓXM plane for the 2% Ce-doped sample—CCMO2 (**a**) and 4% Ce-doped—CCMO4 (**b**) measured with hv at the Mn 2p resonance. The white square designates the cubic pseudo-cell surface Brillouin zone (BZ). The doping clearly increases the experimental Luttinger volume. **c** Out-of-plane FS map for CCMO2 in the ΓZR plane recorded under variation of hv. **d** Sketch of the 643 eV line cutting the $\Gamma_8$ point in the second BZ in $k_\parallel$ and the trajectory of the 643 eV energy in the **k**-space.

| | 3D-electrons | | q2D-electrons | |
|---|---|---|---|---|
| | $k_F$ (Å$^{-1}$) | $n_{3D}$ (cm$^{-3}$) | $k_F$ (Å$^{-1}$) | $n_{q2D}$ (cm$^{-3}$) |
| 2% Ce-doped CCMO2 | 0.10 ± 0.02 | (0.35 ± 0.06) × 10$^{20}$ | 0.13 ± 0.02 | (1.20 ± 0.06) × 10$^{21}$ |
| 4% Ce-doped CCMO4 | 0.20 ± 0.02 | (2.80 ± 0.06) × 10$^{20}$ | 0.15 ± 0.02 | (1.60 ± 0.06) × 10$^{21}$ |

Table 1 Electron density for the three dimensional and quasi-two dimensional bands deduced from angle resolved phtoelectron spectroscopy data.

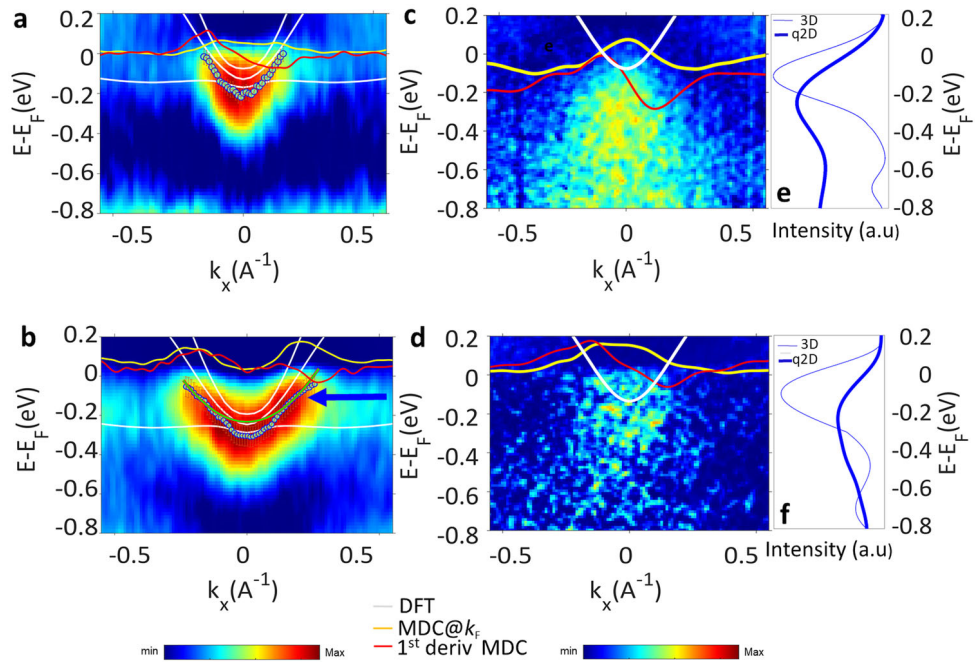

**Fig. 3 Bandstructure of CaCeMnO$_3$. a–d** Band dispersions $E(\mathbf{k})$ measured at the Mn 2p resonance for **a**, **c** 2% Ce-doped CCMO2 and **b**, **d** 4% Ce-doped CCMO4. $E(\mathbf{k})$ of the 3D bands around the Γ-point (**a**, **b**) identifies light electron charge carriers. Blue arrow indicates the threshold energy of energy distribution curves (EDC) maxima deviating in CCMO4 from the parabolic dispersion. Angle resolved photoelectron spectroscopy (ARPES) images of the quasi-2D bands around $k_y = 0.5\pi/a$ (**c**, **d**) show massive humps extending down in binding energy ($E_B$) which manifest heavy polaronic charge carriers. Also shown through (**a**–**d**) are the overlaid density functional theory (DFT)-theoretical bands and gradients of the energy-integrated ARPES intensity, identifying the Fermi wavevector $k_F$.; **e**, **f** Spectral function $A(\mathbf{k},\omega) \propto$ EDC at $k_x = k_F$ for the 3D bands (thin lines) and at $k_y = 0.5\pi/a$ for the q2D ones (thick lines) for CCMO2 (**e**) and CCMO4 (**f**). For the q2D bands, the whole $A(\mathbf{k},\omega)$ is dominated by the polaronic hump.

**Weakly coupled electrons in 3D bands**. We will now analyse the experimental band dispersions $E(\mathbf{k})$ and spectral functions $A(\mathbf{k},\omega)$, and demonstrate that the 3D and q2D bands provide charge carriers having totally different nature. First, we focus on the former derived from the $3z^2 - r^2$ $e_g$-orbitals. Their $k_x$-dispersions along the ΓX-direction of the bulk BZ marked in Fig. 2a, b (cut "A") are visualized by the ARPES intensity images for the CCMO2 (d) and CCMO4 (f) samples (for band structure data through an extended **k**-space region see Supplementary Figs. 4, 5). The ARPES dispersions confirm the picture of doping-dependent band filling. They are shown overlaid with the DFT-calculated bands slightly shifted to match the experimental $k_F$, assuming deviations from the nominal doping of at most 25%[33].

Perovskite oxides are systems where a coupling of the electrons with various bosonic excitations such as plasmons[34] or phonons[35,36] is expected, further translating in band renormalization and modifications of the electron effective mass $m^*$ and mobility. In search of bosonic coupling signatures, we evaluated maxima of the ARPES energy-distribution curves (EDCs), shown in Fig. 3a, b as filled circles. The raw data are presented in Supplementary Fig. 6 and data-processing details are in Supplementary Note 2. For CCMO2, these points follow a parabolic $E(\mathbf{k})$ with $m^* \sim 0.35 m_0$ ($m_0$ is the free-electron mass) in agreement with the DFT predictions. With the increase of occupied bandwidth in CCMO4, the experimental points start deviating from the parabolic dispersion at $E_B' = 80 \pm 20$ meV while approaching $E_F$, and below this energy the experimental dispersion follows the same $m^* \sim 0.35 m_0$. The parabolic fit of the region above $E_B'$ is the green line in Fig. 3b. Such a dispersion discontinuity, or a kink, is a standard signature of weak electron-boson coupling. However, its quantitative analysis in terms of coupling strength suffers from the experimental statistics and possible admixture of the $x^2 - y^2$ $e_g$ band. Similar behavior in the CCMO2 data may be hidden behind smaller occupied bandwidth. The same feature is also identified based on the analysis of momentum-distribution curves (MDCs) in Supplementary Fig. 7.

Which are the bosonic excitations which could manifest as the dispersion kink? One can rule out magnons because these excitations in the parent AFM-ordered CMO strongly couple to phonons[37] but only marginally to electrons. Furthermore, inelastic neutron scattering experiments[38] find the energy of magnons associated with the AFM spin orientations at ~20 meV that is well below the experimental kink energy. The magnons associated with weak FM due to canted AFM spin orientation[3] of our CCMO samples grown under compressive strain will have yet smaller energy because the FM spin coupling is much weaker compared to the AFM one. Other bosonic candidates, plasmons, can also be excluded on the basis of their calculated energies (see Supplementary Fig. 8), which are almost an order of magnitude larger than the observed kink energy. In turn, the calculated energies $\omega_{ph}$ of phonons, which could show significant EPI matrix elements, are found to fall into the 50–80 meV energy range (see Supplementary Fig. 9) matching the experimental data. We therefore assign the kink to EPI. The presence of a kink in the dispersions suggests that the 3D-electrons stay in the weak-coupling regime of EPI, forming a subsystem of light charge carriers.

**Strongly coupled polarons in quasi-2D bands.** We now switch to the $e_g$ $x^2 - y^2$ derived q2D bands forming the FS cylinders. The ARPES images along $k_x$-cuts of these cylinders marked in Fig. 2a, b (Cut "B") are shown in Fig. 3c, d for CCMO2 and CCMO4, respectively, overlaid with the DFT-calculated bands. Surprisingly, the cuts do not reveal any well-defined bands but only humps of ARPES intensity falling down from the $e_g$ $x^2 - y^2$ dispersions. The experimental $k_F$ for CCMO2 and CCMO4, and the corresponding $n_{q2D}$ values compiled in Table 1 reveal that, importantly, $n_{q2D}$ exceeds $n_{3D}$ by roughly an order of magnitude.

Remarkably, the experimental $A(\mathbf{k},\omega)$ at $k_F$, shown in Fig. 3e, f do not exhibit any notable quasi-particle (QP) peak but only a broad intensity hump extending down in $E_B$. Such a spectral shape with vanishing QP weight is characteristic of electrons interacting very strongly with bosonic degrees of freedom, which form heavy charge carriers.

Based on the above analysis of the characteristic phonon frequencies in CCMO, these bosons should be associated with phonons interacting with electrons through strong EPI. Physically, in this case a moving electron drags behind it a strong lattice distortion—in other words, a phonon cloud—that fundamentally reduces mobility of such a compound charge carrier called polaron[39,40]. An overlap of different phonon modes (Supplementary Fig. 9) and their possible dispersion explains the experimentally observed absence of any well-defined structure of the hump. Tracking the occupied part of the q2D band, the hump is however confined in **k**-space. This fact rules out the defect-like small polarons, whose weight spreads out through the whole BZ, and suggests the picture of large polarons associated with a long-range lattice distortion[41]. The simultaneous manifestation of very strong EPI with the large-polaron dispersion points to long-range EPI[41], while the absence of major modifications in the spectral function shape during temperature-dependent measurements (Supplementary Fig. 10) suggests that the phonons involved in the polaronic coupling have frequencies above ~20 meV. The polaron formation in CCMO bears resemblance to the better studied parent CMO[19,22] where EPI is also associated with several phonon modes with frequencies $\omega_{ph}$ in the 50–80 meV range[42]. We note that while these previous works assumed a small polaron picture of the charge carriers in CMO[23,24], our ARPES results suggest their large-polaron character. Our identification of the heavy polaronic charge carriers in the $e_g$ $x^2 - y^2$ bands is also supported by the doping dependence of the ARPES data in Fig. 3c, d: the increase of $n_{q2D}$ from CCMO2 to CCMO4 reduces the hump spectral density and shifts it upwards to the quasiparticle band, which indicates a decrease of the EPI strength. The dichotomy of the 3D and q2D carriers in terms of different

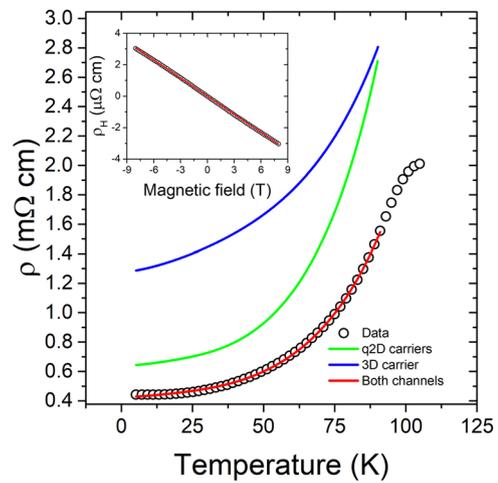

**Fig. 4 Transport properties.** Temperature dependence of resistivity for a 30 nm 4% doped CaMnO$_3$ (CCMO4) sample, and its fit using a two-band model with the light-carrier and heavy-carrier channels plotted together with the individual contributions of the channels. Inset: Hall effect for this sample at 130 K, and its simulation using carrier concentrations $n_e = 2.8 \times 10^{20}$ cm$^{-3}$ and mobility $\mu = 9.9$ cmV$^{-1}$s$^{-1}$ for the light-carrier channel, respectively $n_e = 1.6 \times 10^{21}$ cm$^{-3}$ and $\mu = 3.6$ cmV$^{-1}$s$^{-1}$ for the heavy-carrier one.

EPI strengths, mobility and $m^*$ identified in the ARPES data should also be consistent with a transport model including two types of carriers, with radically different masses and mobilities.

**Light and heavy electrons in transport.** We will now examine how these findings translate into transport experiments performed for CCMO4 (Fig. 4) and show that, indeed, the Hall resistivity is consistent with a two band model combining the contribution of carriers with different effective masses and mobilities. This analysis has only been done for CCMO4 that remains metallic in the whole temperature range below $T_c$. In CCMO2, in contrast, a thermally activated behavior is found at low temperature, already suggesting carrier localization and/or strong temperature dependent enhancement of $m^*$.

The above experimental data is simulated in a two-band model, using carrier densities computed from the ARPES data (see Table 1) for heavy and light charge carriers as follows:

$$\rho^{xy} = \frac{B}{q} \frac{(n_1 \mu_1^2 + n_2 \mu_2^2) + (n_1 + n_2)(\mu_1 \mu_2 B)^2}{(n_1 \mu_1 + n_2 \mu_2)^2 + (n_1 + n_2)^2 (\mu_1 \mu_2 B)^2} \quad (1)$$

with $\rho^{xy}$ the Hall resistivity, $n_i$ the charge density of each conducting channel, $\mu_i$ their corresponding mobilities, $B$ the magnetic field, and $q$ the elementary charge.

The longitudinal resistivity is obtained from the equation:

$$\frac{1}{\rho^{xx}} = \frac{1}{\rho_1^{xx}} + \frac{1}{\rho_2^{xx}}, \quad (2)$$

where $\rho_1^{xx}$ and $\rho_2^{xx}$ are the longitudinal resistivities of the two different carriers, which correspond to two non-interacting channels that conduct in parallel.

For the heavy carriers, the scattering rate is dominated by scattering with impurities, magnons, and phonons in the presence of the strong EPI that gives rise to the formation of polarons[43].

Table 2 Fit parameters for transport data.

| Carriers | $\rho_0$ | $\rho_{1.5}$ | $\rho_5$ | $\rho_{e-ph}$ | $g^2$ |
|---|---|---|---|---|---|
| quasi-2D | 19.1 | 0.015 | | 0.02 | 3.37 |
| 3D | 38.2 | 0.03 | $6.3 \times 10^{-9}$ | | |

Contributions at the total resistivity of impurity scattering ($\rho_0$), magnon scattering ($\rho_{1.5}$), weak electron–phonon scattering ($\rho_5$) and strong electron-phonon interaction ($\rho_{e-ph}$).

The resistivity for the polaronic carriers is:

$$\rho_1^{xx} = \rho_P(T) = \rho_0^p + \rho_{1.5}^p T^{1.5} \\ + \rho_{e-ph} \frac{\omega_0}{\sinh^2(\hbar\omega_0/2k_BT)} \exp[g^2 \coth(\hbar\omega_0/2k_BT)] \quad (3)$$

with $\rho_0$, $\rho_{1.5}$, and $\rho_{e-ph}$, the contributions at the total resistivities resulting from impurity scattering, magnon scattering and respectively phonon scattering and $g^2$ the band narrowing factor.

On the other hand, although the light carriers are subject to the same scattering mechanisms, the EPI is only weak and the scattering rate with phonons follows a $T^5$ power law, due to the Bloch-Gruneisen temperature being much higher than the critical temperature of CCMO, which can be expressed in the form:

$$\rho_2^{xx} = \rho_F(T) = \rho_0^F + \rho_{1.5}^F T^{1.5} + \rho_5 T^5 \quad (4)$$

For details on the derivation of (3) and (4) see Supplementary Note 3.

The results of the fit are in Fig. 4, and the obtained values for the fit parameters are presented in Table 2. The Hall resistivity shows a very weakly non-linear dependence on the magnetic field, consistent with the two types of charge carriers. Using carrier densities computed from the ARPES data (see Table 1) for heavy and light charge carriers, we obtain the correct trend for their mobilities, with the heavy-carrier mobility values smaller by a factor of ~3. This is consistent with larger $m^*$ of these charge carriers inferred from their non-dispersive character which results from strong polaronic coupling found in ARPES, and confirms the derived dichotomy of the heavy electrons strongly coupled with lattice vibration and of the light ones in weakly coupled regime.

## Discussion

Why is the polaronic coupling so strong for the heavy charge carriers, in contrast to the light ones where such effects lead to only weak coupling? At first sight, the difference might be connected with different EPI matrix elements for the corresponding $t_{2g}$ and $e_g$ derived wavefunctions. Surprisingly, our DFT-based EPI calculations have found similar values of these matrix elements (Supplementary Fig. 11 and Supplementary Table 1). Moreover, the difference in wavefunction character does not explain the extreme sensitivity of polaronic coupling of the heavy charge carriers to doping.

The answer should lie in different Fermi energy $\Delta\varepsilon_F$ in each case, i.e., the occupied bandwidth between the band bottom and $E_F$. In the case of the 3D bands, the experiment shows $\Delta\varepsilon_F > \omega_{ph}$, and the EPI manifestations in $A(\mathbf{k},\omega)$ do not go beyond the formation of a kink typical of a weak EPI strength. This situation is described within the conventional Migdal theorem (MT), neglecting high-order vertex corrections to the EPI self-energy. In this case the strong coupling regime cannot be reached and the quasiparticle peak stays clearly discernible. In the case of the q2D bands, on the other hand, we have $\Delta\varepsilon_F \lesssim \omega_{ph}$. The MT is no longer valid in this situation, and high-order vertex corrections to EPI become important. The system enters into the strong coupling regime, as expressed by a broad red-shifted phonon hump of $A(\mathbf{k},\omega)$. Indeed, recent calculations using the exact Diagrammatic Monte Carlo method[44] have demonstrated, at the time without direct experimental verification, that the vertex corrections are indispensable to capture the strong coupling regime of polaronic renormalization and that their suppression just extinguishes the weak to strong coupling crossover. Furthermore, recent calculations[45] within the many-body Bold-Line Diagrammatic Monte Carlo method[46] have shown that these corrections become important when $\Delta\varepsilon_F \approx \omega_{ph}$. This effect has also been suggested for the electron pairing mechanism in cuprates[47,48]. In our case, $\Delta\varepsilon_F$ of the q2D bands falls in the energy range around 50–80 meV of the EPI-active phonons (Supplementary Fig. 11), invoking the strong polaronic coupling through the MT-breakdown and concomitant boost of the vertex corrections. As in CCMO the heavy-polaron $n_{q2D}$ much prevails over the light-electron $n_{3D}$, their large $m^*$ dramatically reduces their transport efficiency at our doping levels. The increase of $\Delta\varepsilon_F$ with band filling explains the weakening of the polaronic effects in these bands when going from CCMO2 to CCMO4 (Fig. 3c, d and Supplementary Fig. 10a, b). We note that the light electrons, fast compared to the lattice oscillations, are much more effective in screening EPI compared to the slow, heavy polarons. Therefore, EPI in the light subsystem is much decoupled from the heavy subsystem. On the other hand, the former can participate in screening of EPI in the latter. Their small $n_{3D}$ compared to $n_{q2D}$ is however insufficient to quench completely the strong EPI for the heavy subsystem. We note that the EPI boost on the verge of MIT can be particularly important in superconductivity, where it can promote phonon-mediated formation of Cooper pairs in materials with otherwise insufficient EPI strength. These ideas cannot however be directly stretched to electron interactions with magnons because their bosonic properties are intrinsically valid only for small bosonic occupation numbers, making vertex corrections small in all circumstances.

Apart from the EPI, essential for the formation of the dichotomic charge carriers in CCMO are electron correlations. Indeed, as evidenced by our DFT + U calculations (Supplementary Fig. 12), they push the q2D bands up in energy, thereby reducing their $\Delta\varepsilon_F$ and triggering the boost of the EPI, much stronger compared to the 3D ones. Therefore, the electron correlations and EPI beyond the MT are two indispensable ingredients of the charge-carrier dichotomy in CCMO. First-principles calculations, incorporating both effects, still remain a challenge.

The EPI boost on the verge of MIT identified here for CCMO should be a hallmark of this transition phenomenon for many TMO systems, including high-temperature superconducting cuprates[47,48]. Indeed, the TMOs are characterized by strong electron coupling with the lattice which, apart from various lattice instabilities such as Jahn-Teller distortions as well as tilting and rotations of the oxygen octahedra[12–15,49,50], gives rise to polaronic activity. Physically, our results show that upon going from the insulating to metallic state of TMOs, electrons do not immediately evolve from localized to delocalized states. Rather, they stay entangled with the lattice, as manifested by the EPI boost, as long as their energy stays on the lattice-excitation energy scale. In our case, tracing of the EPI-induced effects in CCMO has been facilitated, in contrast to other TMOs, by the absence of structural transitions interfering with the phonon excitations and, in contrast to other strongly correlated systems like cuprates and pnictides, by the favorable situation that all elementary excitations other than phonons, including electron interactions, magnons and plasmons, have much different energy scales. The MT-breakdown effects should not however be pronounced in conventional semiconductors like Si or GaAs, where relatively weak electron coupling to the lattice at all doping levels is evidenced, for example, by temperature-independent lattice structure up to the melting point. Finally, by identifying the interplay between electron correlations and EPI as the origin of the $m^*$ enhancement in lightly-doped CMO, our work sheds light on the recent observation of a giant topological Hall effect in this material[21], and suggests strategies for its realization in other systems. To the best of our knowledge, our study on CCMO is the first case where a dramatic distinction of charge carriers has been unambiguously identified as tracing back not to merely one-electron band structure or electron-correlation effects like in pnictides, for example, but to peculiarities of electron-boson coupling.

Summarizing, our combined SX-ARPES and transport study supported by DFT-based calculations of EPI in Ce-doped CMO has established the co-existence of two dichotomic charge-carrier subsystems dramatically different on their EPI-activity: (1) light electrons forming a Fermi liquid in the 3D $e_g$ $3z^2 - r^2$ derived bands, where the weak EPI could manifest as band dispersion kinks, and (2) heavy large polarons in the q2D $e_g$ $x^2 - y^2$ bands, where correlations reduce the occupied bandwidth $\Delta\varepsilon_F$ to be comparable or smaller than $\omega_{ph}$, pushing the system into a regime where the MT breaks down. This invokes strong vertex corrections to the electron self-energy, and EPI boosts with an almost complete transfer of the spectral weight to the polaronic hump. The increase of doping in CCMO progressively increases $\Delta\varepsilon_F$ of the q2D bands and thus, by the MT recovery and concomitant reduction of EPI, the mobility of the corresponding heavy charge carriers. An alternative route to tune EPI in CCMO is compressive or tensile epitaxial strain which alters filling and thus $\Delta\varepsilon_F$ of the 3D vs. q2D bands[51,52]. Our findings disclose the previously overlooked important role the EPI can play in MITs even caused by purely electronic mechanisms.

## Methods

**Sample growth.** 20 nm and 30 nm $Ca_{1-x}Ce_xMnO_3$ ($x$ = 2%, 4% nominal Ce concentrations) thin films were grown by pulsed laser deposition from stoichiometric targets on (001) $YAlO_3$ substrates using a Nd:YAG laser. Commercial $YAlO_3$ (001) oriented substrates were prepared with acetone cleaning and ultrasound in propanol, and then annealed at 1000 °C in high $O_2$ pressure. The substrate temperature ($T_{sub}$) and oxygen pressure ($P_{O_2}$) during the deposition were 620 °C and 20 Pa, respectively. Post-deposition annealing was performed at $T_{sub} \approx 580$ °C and $P_{O_2}$ = 30 kPa for 30 min, followed by a cool-down at the sample oxygen pressure. The thickness of the $Ca_{1-x}Ce_xMnO_3$ thin films was measured by X-ray reflectivity with a Bruker D8 DISCOVER. The samples are ~1% compressively strained at the $YAlO_3$ (001) in-plane lattice constant[21,33]. Magnetotransport measurements were performed on 30 nm thick samples using measurement bridges patterned by optical lithography and Ar ion etching. Electrical contacts for measurements were made on platinum electrodes defined by a combination of lithography and lift-off techniques.

**SX-ARPES experiment.** The virtues of SX-ARPES include the increase of photoelectron mean free path $\lambda$ with energy and the concomitant reduction of the intrinsic broadening $\Delta k_z$, defined, by the Heisenberg uncertainty principle, as $\Delta k_z = \lambda^{-1}$[27,28]. Combined with free-electron dispersion of high-energy final states, this allows sharp $k_z$-resolution and thereby accurate navigation in **k**-space of 3D materials like CCMO[20,28]. SX-ARPES experiments were carried out at ADRESS beamline at Swiss Light Source which delivers high soft-X-ray photon flux of more than $10^{13}$ photons$^{-1}$ s$^{-1}$ 0.01% bandwidth. The endstation[29], operating at a grazing X-ray incidence angle of 20°, used the analyzer PHOIBOS-150 (SPECS GmbH). The combined (beamline+analyzer) energy resolution was about 65 meV at 643 eV. The measurements were performed on 20 nm thick samples, at low temperature $T$ = 40 K using circularly polarized X-rays. In order to prevent sample degrading under the X-ray beam and the accompanying reduction of Ce from $Ce^{4+}$ to $Ce^{3+}$ and of $Mn^{4+}$ to $Mn^{3+}$ (Supplementary Figs. 11, 12) SX-ARPES measurements were conducted in oxygen atmosphere at a partial pressure of $6.9 \times 10^{-7}$ mbar. Photoelectron kinetic energy and emission angle were converted to **k** with a correction for photon momentum $hv/c$, where $c$ is the speed of light[29]. The value of the inner potential used to render the incoming energy $hv$ to $k_z$ was $V_0$ = 10 eV[53]. The dispersive structures in the ARPES images were emphasized by subtracting the non-dispersive spectral component obtained by angle integration of the raw data (Supplementary Fig. 9).

**First-principles calculations.** First-principles calculations were carried out for CMO and Ce-doped CMO in the orthorhombic $Pnma$ structure with 20 atoms in the primitive unit cell, using spin-polarized density-functional theory (DFT) as implemented in the *Quantum Espresso* package[54,55]. The core-valence interaction was described by means of ultrasoft pseudopotentials, with the semicore 3s and 3p states taken explicitly into account in the case of Ca and Mn. The calculations were converged by using a plane-wave cutoff of 60 Ry and a $6 \times 4 \times 6$ Brillouin-zone (BZ) grid for the antiferromagnetic (AFM) ground state with G-type order. Doping was described within the rigid-band approximation using the nominal Ce concentrations of 2 and 4%. The 2% doping due to the coexisting $Mn^{4+}/Mn^{2+}$ states corresponds to additional 0.16 e/unit cell, and 0.32 e/unit cell for 4% Ce. Charge neutrality was maintained by including a compensating positively-charged background. A denser $10 \times 8 \times 10$ BZ grid was used to converge the ground-state properties in the doped case.

**Electrical and magnetic characterization.** The magnetotransport characterization of the CCMO samples was performed in a Quantum Design Physical Properties Measurement System (PPMS) Dynacool. The temperature dependence of the resistivity was measured at a constant current of 5 μA during a warming run after field cooling. For Hall measurements, magnetic fields were swept up to ±8 T. To separate the Hall contribution from that of the longitudinal magnetoresistance, an antisymmetrization procedure was performed by separating the positive and negative field sweep branches, interpolating the two onto the same field coordinates and then antisymmetrizing using:

$$\rho'_\pm(+H) = [\rho_\pm(+H) - \rho_\mp(-H)]/2$$

## Data availability

The data presented are available from the corresponding authors upon reasonable request

## Acknowledgements

This research received financial support from the ERC Consolidator grant "MINT" (contract number n°615759), ANR project "FERROMON" and a public grant overseen by the ANR as part of the "Investissement d'Avenir" program (LABEX NanoSaclay, ref. ANR-10-LABX-0035) through projects "FERROMOTT" and "AXION". A.S.M. acknowledge the support of the ImPACT Program of the Council for Science, Technology and Innovation (Cabinet Office, Government of Japan) and JST CREST Grant Number JPMJCR1874, Japan. C.V. and F.G. acknowledge support from the Leverhulme Trust (Grant RL-2012-001), the Graphene Flagship (Horizon 2020 Grant No. 785219—GrapheneCore2) and the UK Engineering and Physical Sciences Research Council (Grant No. EP/M020517/1). C.V. and F.G. also acknowledge the use of the University of Oxford Advanced Research Computing (ARC) facility, the ARCHER UK National Supercomputing Service under the "T-Dops" project, the DECI resource "Cartesius" based in The Netherlands at SURFsara and "Abel" based in Oslo with support from the PRACE AISBL, PRACE for awarding us access to MareNostrum at BSC-CNS, Spain, and CSD3, UK EPSRC (Grant EP/P020259/1). F.B. acknowledges funding from Swiss National Science Foundation under Grant Agreement No. 200021_146890 and from European Community's Seventh Framework Programme (FP7/2007-2013) under Grant Agreement No. 290605 (PSIFELLOW/COFUND). M.-A.H. acknowledges the support by the Swiss Excellence Scholarship grant ESKAS-no. 2015.0257 and by the NIMP Core Program 21N2019 from Romanian Ministry of Research and Innovation.

## Author contributions

V.N.S. and M.B. conceived the combined SX-ARPES and transport study of CCMO. M.-A.H., V.N.S., F.B., L.L.L., L.V, V.G., J.R., and A.S. performed the SX-ARPES experiment supported by M.B. and T.S. M-A.H. and V.N.S. processed and interpreted the SX-ARPES data in terms of dichotomic charge carriers. L.V. and A.S. assisted by V.G. and M.B. manufactured the samples, and L.V performed their structural and Hall characterization. C.V., F.G., and M.-A.H. performed the DFT calculations. C.V. and F.G performed the DFPT and RPA calculations. A.S.M. formulated the MT-breakdown concept. A.S.M., C.V., and F. G. elaborated the physics of EPI. M.-A.H. and V.N.S. wrote the manuscript with contributions of L.V., A.S.M., C.V., M.B., and F.G. All authors discussed the results, interpretations, and scientific concepts.

## Competing interests

The authors declare no competing interests.

## Additional information

**Supplementary information** is available for this paper at https://doi.org/10.1038/s42005-020-0330-6.

**Correspondence** and requests for materials should be addressed to M.-A.H. or V.N.S.

**Reprints and permission information** is available at http://www.nature.com/reprints

**Publisher's note** Springer Nature remains neutral with regard to jurisdictional claims in published maps and institutional affiliations.

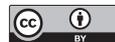